\documentclass[conference,10pt]{IEEEtran}
\IEEEoverridecommandlockouts
\usepackage{cite}
\usepackage{hyperref}
\usepackage{amsmath,amssymb,amsfonts, color,bm}
\usepackage{algorithmic}
\usepackage{graphicx}
\usepackage{textcomp}
\usepackage{xcolor}
\usepackage{enumitem}
\usepackage[linesnumbered,ruled,vlined]{algorithm2e}

\ifCLASSOPTIONcompsoc
\usepackage[caption=false, font=normalsize, labelfont=sf, textfont=sf]{subfig}
\else
\usepackage[caption=false, font=footnotesize]{subfig}
\fi

\def\BibTeX{{\rm B\kern-.05em{\sc i\kern-.025em b}\kern-.08em
    T\kern-.1667em\lower.7ex\hbox{E}\kern-.125emX}}
\def\x{{\boldsymbol x}}
\def\y{{\boldsymbol y}}
\def\z{{\boldsymbol z}}

\begin{document}

\title{Variational Speech Waveform Compression to Catalyze Semantic Communications
\thanks{This work was supported in part by the National Natural Science Foundation of China under Grant 92067202, Grant 62001049, and Grant 62071058, in part by the Beijing Natural Science Foundation under Grant 4222012. \emph{(Corresponding author: Jincheng Dai, email: daijincheng@bupt.edu.cn)}}
}
\author{\IEEEauthorblockN{Shengshi~Yao\IEEEauthorrefmark{1}, Zixuan~Xiao\IEEEauthorrefmark{1}, Sixian~Wang\IEEEauthorrefmark{1}, Jincheng~Dai\IEEEauthorrefmark{1}, Kai~Niu\IEEEauthorrefmark{1}\IEEEauthorrefmark{2}, and Ping~Zhang\IEEEauthorrefmark{1}}
\IEEEauthorblockA{\IEEEauthorrefmark{1}Beijing University of Posts and Telecommunications, Beijing, China\\
\IEEEauthorrefmark{2}Peng Cheng Laboratory, Shenzhen, China}
}

\maketitle

\begin{abstract}
We propose a novel neural waveform compression method to catalyze emerging speech semantic communications. By introducing nonlinear transform and variational modeling, we effectively capture the dependencies within speech frames and estimate the probabilistic distribution of the speech feature more accurately, giving rise to better compression performance. In particular, the speech signals are analyzed and synthesized by a pair of nonlinear transforms, yielding latent features. An entropy model with hyperprior is built to capture the probabilistic distribution of latent features, followed with quantization and entropy coding. The proposed waveform codec can be optimized flexibly towards arbitrary rate, and the other appealing feature is that it can be easily optimized for any differentiable loss function, including perceptual loss used in semantic communications. To further improve the fidelity, we incorporate residual coding to mitigate the degradation arising from quantization distortion at the latent space. Results indicate that achieving the same performance, the proposed method saves up to 27\% coding rate than widely used adaptive multi-rate wideband (AMR-WB) codec as well as emerging neural waveform coding methods.
\end{abstract}

\section{Introduction}
\label{sec:introduction}
Waveform coding and parametric coding are two mainstream categories of speech coding methods. Waveform coding aims to produce a high-fidelity reconstruction with a decent compression ratio for efficient transmission in a communication system. Parametric codec introduces a parametric decoder that synthesizes speech from sets of acoustic features, where the features are encoded and compressed as conditional variables for the decoder. In this work, we consider speech waveform coding based on artificial neural networks (ANN).

The existing neural waveform coding methods are characterized by an auto-encoder combined with a trainable quantization module \cite{agustsson2017soft, oord2017neural}. Due to the advances in deep learning, vector quantization (VQ) has been applied to ANN-based speech coding to compress the latent feature of speech more efficiently \cite{oord2017neural, theis2017lossy}. It is optimal when the best reproduction codebook is found, and the theoretical limits of performance of VQ have been investigated \cite{vqspeech}. Despite its optimality, it's computationally expensive in VQ as the size of the codebook is increasing exponentially with the rate. Vector quantized variational auto-encoders \cite{oord2017neural} (VQ-VAE) concentrates on discrete latent representation through a parametrization of the posterior distribution of discrete latents. It performs well on a low latent capacity (the dimensionality of the latent space) in parametric codecs. VQ-VAE followed with a WaveNet \cite{kleijn2018wavenet} generative decoder achieves a low coding rate at 1.6 kbps ($10^3$ bit per second), yet at the cost of complexity from the generative model \cite{garbacea2019low}. The generative decoder does not ensure a faithful reconstruction of raw audio. However, on the condition of a large latent capacity, which implies a high bitrate, the size of the VQ codebook and the complexity of searching the codebook soar.

Another critical feature of speech codec is rate scalability. VQ-VAE itself does not support rate scalability intrinsically, where the prior is assumed constant and uniform \cite{oord2017neural}. To enable rate control, existing works \cite{kankanahalli2018end, zhen2019cascaded, zhen2020efficient} imposed a constraint on the entropy of quantized latent features to formulate the rate-distortion (RD) objective. However, due to the high complexity of VQ mentioned previously, the entropy is estimated over the marginal distribution of the scalar quantization bins. Thus, the dependency among the latent features is ignored, i.e., the actual distribution of latent features is not well captured.

Inspired by neural image compression \cite{balle2016end, balle2018variational, balle2020nonlinear}, we further consider the dependency within the latent features of speech frames rather than quantizing and encoding them directly. Despite linear models in traditional digital signal processing, the nonlinearity and linearity of raw waveform are jointly analyzed, yielding latent features. In particular, the dependencies of the latent speech features are well learned by a pair of hyperprior transform, and accordingly, an entropy model is established to guide the entropy coding. To catalyze emerging semantic communications \cite{zhang2022toward, weng2021semantic, dai2022communication, dai2022nonlinear, niu2022paradigm}, the target of speech compression is not limited to waveform fidelity. Perceptual metrics are considered in model training to align with human-to-human semantic communications. Both waveform distortion and perceptual distortion are included in the RD Lagrangian objective to achieve a rate-perception-distortion trade-off. On that basis, we investigate a residual coding scheme to mitigate the degradation arising from quantization, which takes place in the latent space.

We verify the performance of the proposed speech waveform coding method across various rates. We observe a noticeable improvement in objective quality scores, compared to existing neural waveform coding methods as well as widely used adaptive multi-rate wideband (AMR-WB) codec \cite{bessette2002adaptive} and Opus codec \cite{valin2012definition}. In addition, adding a residual latent feature coding branch further reduces the impact of quantization of latent features and shows substantial gain at high bitrates. With the decrease in the target rate, the residual branch tends to encode the residuals with fewer bits. Achieving the same target of objective quality scores, our method can save up to $27\%$ coding rates in low bitrate region compared to traditional codecs.

\section{Preliminaries}
\label{sec:preliminaries}

\subsection{Vector Quantization in Speech Coding}

Vector quantization (VQ) is a crucial technique to achieve high compression performance in both traditional and neural source coding. It is optimal when the correlation of the speech signal is decomposed completely. VQ was first introduced to speech coding in linear predictive coding (LPC) since when VQ is widely used in speech and audio codecs. Researchers have been devoted to reducing the computational and storage cost of VQ and codebook search algorithms \cite{vqspeech}. In neural speech coding, vector quantized variational auto-encoder (VQ-VAE) \cite{oord2017neural} replaces continuous latent vectors with codebook vectors based on the nearest distance rule.

\subsection{Residual Coding in Speech Coding}

Residual coding, a well-formalized technique applied in compressing multi-media signals, is to encode the difference between the actual input and what the codec predicts. \cite{zhen2019cascaded} encodes raw signals or residual signals (error of linear prediction) in cascaded stages where each one encodes what is not reconstructed from preceding modules. However, this multi-stage solution is auto-regressive, i.e., the module input relies on the output of the preceding one. It becomes intractable in communication systems. SoundStream \cite{zeghidour2021soundstream} cascades layers of VQ performing quantization iteratively to reduce the codebook size, yet with equal rate allocation for each stage. In this work, we propose a residual branch to encode the residual of latent arising from quantization as a compensation, and the rate allocation for the primary latent and the residual of latent is well-tuned by setting a weighted RD objective.

%
%

\section{Method}
\label{sec:NTC waveform codec}
In this section, we elaborate on the proposed speech waveform compression method based on nonlinear transform. Based on this backbone, we propose a novel residual coding method. Finally, the training techniques are introduced.

\subsection{Architecture}
We consider wideband signal input with a sampling rate of 16 kHz. The waveform is firstly processed to a stack of frames $\x\in \mathbb{R}^{N\times C\times L}$, where $N$ is the number of frames, $C$ is the number of sound channels, and $L$ is the frame length (the number of sampling point). Its probability is given as $p_{\x}(\x)$.

\begin{figure}[htbp]
	\centering
	\includegraphics[scale=0.81]{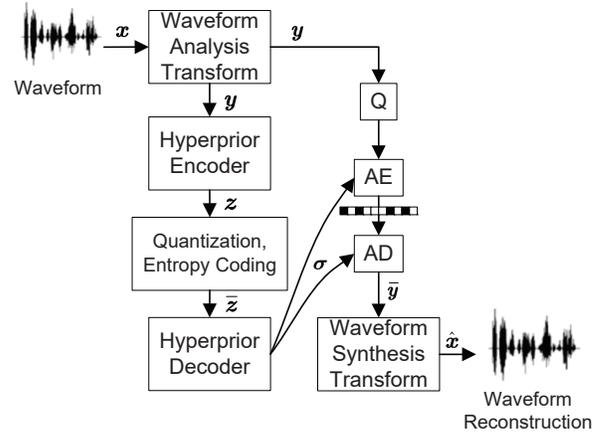}
	\caption{Network architecture of the proposed model of speech waveform coding. In the training phase, $\bar{\y}, \bar{\z}$ are replaced by $\tilde{\y}, \tilde{\z}$, respectively.}
	\label{fig1_network}
\end{figure}

Fig. \ref{fig1_network} illustrates the overall architecture of our proposed method. The waveform analysis transform, parameterized by $\bm{\phi}_g$ is a stack of convolutional neural networks, transforms frames $\x$ into latent feature $\y = g_{a,\bm{\phi}_g}(\x)$. Different strides and dilations are used in order to capture various resolutions of speech features. The dilated convolution block is composed by layers with increasing dilations, from coarse to fine. Inspired by ResNet \cite{he2016deep}, each convolutional block is connected by a shortcut. At the decoder side, given reconstructed latent feature $\hat{\y}$, the synthesis transform recovers the speech signals as $\hat{\x} = g_{s,\bm{\psi}_g}(\bar{\y})$, where $\bm{\psi}_g$ encapsulates the neural network parameter of the synthesis transform.

For compression purpose, the latent feature $\y$ is to be quantized as $\bar{\y}$ and then entropy coded. Then, the expected length (rate $R$) of the compressed bit sequence is equal to the entropy of $\bar{\y}$, i.e.,
\begin{equation}
R=\mathbb{E}_{\x \sim p_{\x}}\left[-\log p_{\bar{\y}}\left(\bar{\y}=Q\left(\y\right)\right)\right].
\end{equation}
As is shown in Fig. \ref{fig1_network}, the latent representation of speech $\y$ is quantized using a uniform scalar quantizer $Q$, rounding to nearest integers $\bar{\y}$. The arithmetic encoder (AE) and decoder (AD) act as the entropy coder. To allow optimization via gradient descent in the model training phase, following \cite{balle2016end}, an offset $\bm{o}$ is added to the speech feature, yielding a proxy quantized representation $\tilde{\y} = \y + \bm{o} = g_{a,\bm{\phi}_g}(\x) + \bm{o}$, where $\bm{o}$ is randomly sampled from a uniform $\mathcal{U}(-\frac{1}{2}, \frac{1}{2}) $. The proxy quantized feature satisfies $p_{\bar{y}}\left( k \right)= p_{\tilde{y}}\left( k \right), \forall k\in \mathbb{Z}$.

With respect to quantization, it is optimal to use VQ and search a codebook, yet leading to high complexity in time and space. Previous work adopted scalar quantization to reduce the codebook size and then estimated the marginal distribution of latent feature for each $y_i \in \y$, $i=1,2,\cdots$. However, the real distribution of $\y$ is not accurately modeled, considering the dependency among its elements, i.e., the entropy of $\y$ is overestimated. Hence, we build an additional set of latent variables $\tilde{\z}$ to represent the dependencies by applying another analysis transform by $\z=h_{a,\bm{\phi}_h}(\y)$, $\z$ is named \emph{hyperprior} of the speech feature. Similarly, in the training phase, $\tilde{\bm{z}}= \bm{z} + \bm{o}$ is a replacement for rounding operation $\bar{\z}=Q(\z)$. We variationally model the proxy quantized latent $\tilde{\y}$ as a multivariate Gaussian. The standard deviation of $\tilde{y}_i$ is predicted as $\sigma_i$ by a hyperprior decoder parameterized by $\bm{\psi}_h$, acting as side information for the entropy coding of $\bar{\y}$. To save the cost of side information, we assumed that $\tilde{y}_i$ is zero-mean and then the posterior distribution of $\tilde{\y}$ given $\tilde{\z}$ is derived as
\begin{equation}
p\left( \tilde{\y}|\tilde{\z};\bm{\psi} _h \right) =\prod_i{\left( \mathcal{N}\left( 0,\sigma_{i}^{2} \right) *\mathcal{U}\left( -\frac{1}{2},+\frac{1}{2} \right) \right)},
\end{equation}
with $\bm{\sigma}=\left [ \sigma_1, \sigma_2, \cdots,\sigma_i,\cdots\right]=h_{s,\bm{\psi}_h}(\tilde{\z})$, $\mathcal{N}\left( 0,\sigma_{i}^{2} \right)$ denoting the zero-mean Gaussian distribution with standard deviation $\sigma_{i}$.
As no prior belief about $\tilde{\z}$ exists, it is modeled by a non-parametric factorized density model \cite{balle2016end}.

To sum up, since the true posterior $p_{\tilde{\y},\tilde{\z}|\x}$ is intractable, we approximate it with a parametric variational density
\begin{equation}
q_{\tilde{\y},\tilde{\z}|\x} = \prod_i{\mathcal{U}(\tilde{y}_i|y_i-\frac{1}{2},y_i+\frac{1}{2})} \prod_j{\mathcal{U}(\tilde{z}_j|z_j-\frac{1}{2},z_j+\frac{1}{2})},
\end{equation}
where $\mathcal{U}(m-\frac{1}{2},m+\frac{1}{2})$ denotes the uniform distribution centered on $m \in \mathbb{R}$ with range from $m- \frac{1}{2}$ to $m+ \frac{1}{2}$. The training objective is to optimize $\bm{\phi}_g, \bm{\psi}_g, \bm{\phi}_h, \bm{\psi}_h$  to minimize the Kullback-Leibler (KL) divergence between the variational density $q_{\tilde{\y},\tilde{\z}| \bm{x}}$ and the true posterior $p_{\tilde{\y},\tilde{\z}| \x}$ over the source $\x$, i.e.,
\begin{equation}
\label{eq_ntc_rd}
\begin{aligned}
&\mathbb{E}_{\bm{x}}D_{\text{KL}}\left[ q_{\tilde{\y},\tilde{\z}|\x}\|p_{\tilde{\y},\tilde{\z}|\x} \right] =\mathbb{E}_{\x}\mathbb{E}_{\tilde{\y},\tilde{\z}\sim q_{\tilde{\y},\tilde{\z}|\x}}[ -\log p\left( \tilde{\y} | \tilde{\z} \right)
\\
&- \log p\left( \tilde{\z}\right) -\log p\left( \x|\tilde{\y} \right) +\log q\left(\tilde{\y},\tilde{\z}|\x \right)] +\text{const},
\end{aligned}
\end{equation}
where the fourth term is also constant because of the constant width of the uniform distribution. The first two terms in \eqref{eq_ntc_rd} denote the rate of encoding latent features and side information, respectively. The third term measures distortion. Hence, it gives rise to an RD optimization problem, where a higher rate allows for lower distortion. The training detail is introduced in subsection \ref{subsec:train}.

The network configuration is listed in Table \ref{tab_arch}. The synthesis transform decoder shares a similar design to the waveform analysis transform encoder with a mirrored design, which is omitted in the table.

\begin{table}[htbp]
\renewcommand{\arraystretch}{1.2}
\caption{Network configuration of waveform analysis transform encoder \& hyperprior encoder using convolutional neural networks.}
\begin{center}
\begin{tabular}{|c|c|c|c|c|}
\hline
\textbf{Module} & \textbf{\#Channel} & \textbf{Kernel}& \textbf{Dilation}& \textbf{Scaling} \\
\hline
\multicolumn{5}{|c|}{Encoder}\\
\hline
Input Conv& 64 & 9 & 0 &  -\\
\hline
Dilated Conv Block $\times 4$&64 & 9 & 1,2,4,8 & -\\
\hline
Downsampling & 64 & 9 & 0 &  2\\
\hline
Dilated Conv Block $\times 4$&64 & 5 & 1,2,4,8 & -\\
\hline
Downsampling & 4 & 5 & 0 &  2\\
\hline
\multicolumn{5}{|c|}{Hyperprior Encoder}\\
\hline
Input Conv& 32 & 9 & 0 &  -\\
\hline
Dilated Conv Block $\times 3$&32 & 9 & 1,2,4& -\\
\hline
Downsampling & 32 & 9 & 0 &  2\\
\hline
Dilated Conv Block $\times 2$&32 & 5 & 1,2& -\\
\hline
Downsampling & 2 & 5 & 0 &  2\\
\hline
\end{tabular}
\label{tab_arch}
\end{center}
\end{table}

\subsection{Residual Coding}
Based on the above backbone, we additionally introduce residual coding to mitigate the quantization loss of latent features. Previous residual coding solutions \cite{zhen2019cascaded} adopted a sequential coding manner where the subsequent coding relies on the reconstruction in previous stages, which is inefficient and impractical for communication systems. We integrate differential coding into the backbone, where the residual coding takes place in the latent space instead of re-encoding the residual waveform. Thus, the residual branch is trained jointly without depending on the quality of the reconstruction.

\begin{figure}[htbp]
	\centering
	\includegraphics[scale=0.9 ]{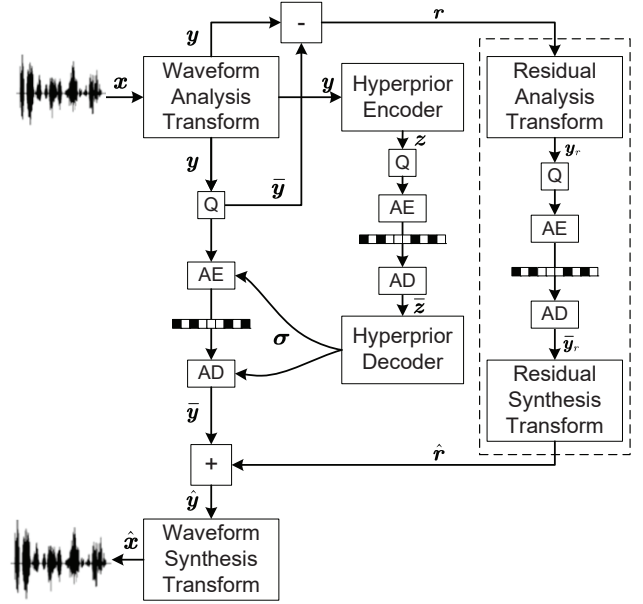}
	\caption{Architecture of speech waveform coding with residual coding of latent features. In the training phase, $\bar{\y}, \bar{\z}, \bar{\y}_r$ are replaced by $\tilde{\y}, \tilde{\z}, \tilde{\y}_r$, respectively.}
	\label{fig2_network}
\end{figure}

The residual branch at the most right of Fig. \ref{fig2_network} encodes the residual latent $\bm{r} = \y - \bar{\y} =\y - Q(\y)$, where $Q(\cdot)$ is the rounding operation. Likewise, $\bm{r}$ is encoded as $\y _r$ by a residual analysis transform. Similar to hyperprior $\z$, $\y _r$ is then modeled by a non-parametric factorized density model, quantized and entropy coded. The inference model recovers the synthesized residuals $\bm{\hat{r}}$ parameterized by a residual synthesis transform. Finally, the waveform synthesis transform reconstructs the waveform signals by merging the primary latent feature $\bar{\y}$ and the residual $\bm{\hat{r}}$, written as $\hat{\x}=g_{s,\bm{\psi}_g}\left( \bar{\y}, \bm{\hat{r}}\right)$.

In our experiments, lightweight convolutional networks are applied as the residual analysis transform and synthesis transform to reduce the extra model complexity. The detail of the training strategy is introduced in the following subsection.

\subsection{Model Training}\label{subsec:train}
As illustrated in \eqref{eq_ntc_rd}, the goal is to optimize the parameter set to achieve a trade-off between the compression ratio and the speech reconstruction quality, i.e., an RD trade-off. Specifically, the loss function $\mathcal{L}$ is written as
\begin{equation}\label{ntc_objective}
  \mathcal{L} =\mathbb{E}_{\x}\left[ -\log p_{\bar{\y}|\bar{\z}}\left( \bar{\y}|\bar{\z} \right) - \log p_{\bar{\z}}\left( \bar{\z} \right) +\lambda d\left( \x,\hat{\x} \right)\right],
\end{equation}
where the Lagrange multiplier $\lambda$ governs the trade-off. The first two terms work out to be the coding rate of the proxy quantized version of $\y$ and $\z$. The third term $d\left( \cdot ,\cdot \right)$ is the distortion between the original waveform and reconstructed one and we adopt euclidian distance.


In the residual coding scheme, additional RD constraint on the residual latent features is required. As the residual $\bm{r}$ of $\y$ is not derivative, the optimization of waveform analysis transform is independent of the residual one. The extra objective function is to optimize the residual analysis/synthesis transform to minimize the KL divergence between $q_{\tilde{\y}_r| \bm{r}}$ and the true posterior $p_{\tilde{\y}_r| \bm{r}}$, i.e.,
\begin{equation}
\label{eq_ntc_res_rd}
\begin{aligned}
&\mathbb{E}_{\bm{r}}D_{\text{KL}}\left[ q_{\tilde{\y}_r|\bm{r}}\|p_{\tilde{\y}_r|\bm{r}} \right] =\mathbb{E}_{\bm{r}}\mathbb{E}_{\tilde{\y}_r \sim q_{\tilde{\y}_r|\bm{r}}}[ -\log p\left( \tilde{\y}_r \right)
\\
& -\log p\left( \bm{r}|\tilde{\y}_r \right)] +\text{const}.
\end{aligned}
\end{equation}

In joint training, the rate of the primary and the residual latent are combined where a seesaw effect exists with different rate allocation strategies for the two branches. Specifically, the rate term is the summation of the rate of three bitstreams in Fig. \ref{fig2_network} with $R=\mathbb{E}_{\x}\left[ -\log _2p\left( \tilde{\y} |\tilde{\z} \right) -\log _2p\left( \tilde{\z} \right) -\log _2p\left( \tilde{\y}_r \right) \right]$ in bit. The result of rate allocation is provided in Section \ref{sec:experiments}.

With regard to the distortion term $d$, it can be designed flexibly to satisfy specific purposes. In human-to-human semantic communications, the waveform fidelity is not all that essential, but the human perceptual similarity between the reconstructed speech waveform and the original one is also important.

Hence, we both count the waveform distortion in time domain, as well as the perceptual loss defined in frequency domain. Mean square error (MSE) between the raw waveform $\x$ and the reconstructed waveform $\hat{\x}$ evaluates the reconstruction error in time domain, formulated as
\begin{equation}
  \mathcal{L}_{\mathrm{MSE}}=\mathbb{E}_{\x}\left\| \x-\hat{\x} \right\|_{2}^{2}.
\end{equation}
As subjective perceptual scores or objective scores usually cannot be directly optimized because of non-differentiable property. Perceptual quality of reconstructed speech waveform is considered in frequency domain. Specifically, we optimize the model to reduce the reconstruction error of mel frequency cepstral coefficient (MFCC) \cite{muda2010voice} features to persue the consistency of the feature in frequency (mel scale) domain with
\begin{equation}
     \mathcal{L}_{\mathrm{perc}}=\mathbb{E}_{\x}\sum_{k=1}^K{\left\| m_k\left( \x \right) -m_k\left( \hat{\x} \right) \right\| _{2}^{2}},
\end{equation}
where $m_k$ is the MFCC function of $k$-th filterbank. Specifically, we choose $K=4$ filterbanks with scale from 8 to 128.

While training the residual coding branch, an additional MSE term $\mathcal{L}_{\mathrm{resMSE}}$ is used to optimize the residual encoder and decoder only, where
\begin{equation}
  \mathcal{L}_{\mathrm{resMSE}}=\mathbb{E}_{\boldsymbol{r}}\left\| \boldsymbol{r}-\boldsymbol{\hat{r}} \right\|_{2}^{2}.
\end{equation}

In a nutshell, the overall loss function to train the model with residual coding is a weighted sum of the loss components as
\begin{equation}\label{eq_loss}
  \mathcal{L}=R+\lambda_{\mathrm{MSE}} \mathcal{L}_{\mathrm{MSE}} + \lambda_{\mathrm{res}} \mathcal{L}_{\mathrm{resMSE}} + \lambda_{\mathrm{perc}} \mathcal{L}_{\mathrm{perc}},
\end{equation}
where the $\lambda$ hyperparameter set controls the ratio of the perception and distortion term with the rate term as reference. It gives rise to a rate-perception-distortion trade-off, and a higher $\lambda$ prompts the model to learn latent representation with higher entropy to reduce the distortion or improve the perceptual quality. In this way, a wide range of bitrates is achievable for our speech coding method.

\section{Experiments}
\label{sec:experiments}
In this section, we provide illustrative numerical results to evaluate the compression and quality of speech waveform coding. Objective quality metric is evaluated, and a subjective listening test is presented to validate our designed approach.

\subsection{Experimental Settings}
The waveform with a single audio channel $C=1$ is sampled at 16 kHz from TIMIT dataset \cite{TIMIT}. The training set contains 3.1 hours of speech from 462 speakers, while the test set contains 0.8 hours of speech. Each frame has $L=512$ samples with an overlap of 32 samples.

\subsection{Results}
For reference, uncompressed wideband speech with 16-bit width has a rate of 256 kbps. As discussed in Section \ref{subsec:train}, it is convenient to tune the coding rate by adjusting the weight of each term in the rate-perception-distortion objective in \eqref{eq_loss}. Multiple coding rates are considered from $\sim$8 kbps to $\sim$24 kbps.

In terms of objective quality evaluation, we report MOS-LQO scores computed from perceptual evaluation of speech quality (PESQ) \cite{pesq} scores, and MOS-LQO score ranges from 1.0 to 4.5. The following methods of neural waveform coding are provided for comparison. Cascaded cross-module residual learning (CMRL) \cite{zhen2019cascaded} encodes the signals in cascaded stages, each of which reconstructs the residual from its preceding modules. ``LPC-CMRL'' introduces a strong prior with an LPC model within a single frame, and then encodes the residual waveform with linear prediction coefficients as an add-on. ``Raw-CMRL'' denotes the circumstance that raw waveform instead of the residual waveform of LPC is compressed by CMRL. CMRL with collaborative quantization (CQ) learns the bit allocation between the LPC coefficients and the residuals \cite{zhen2020efficient}. AMR-WB \cite{bessette2002adaptive} and Opus \cite{valin2012definition} are provided as the representatives of traditional speech codecs. AMR-WB bitrates range from 6 kbps through 24 kbps, yet with predetermined options.

\begin{figure}[htbp]
	\centering
	\includegraphics[scale=0.5]{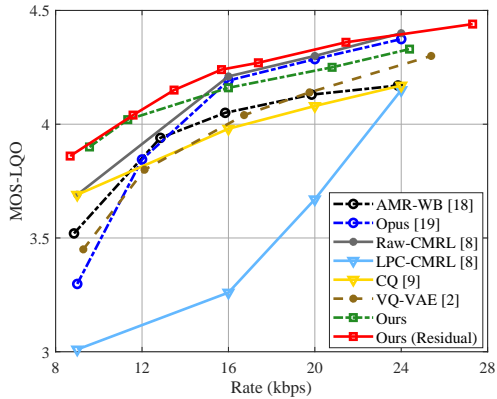}
	\caption{Comparison of MOS-LQO scores for speech waveform coding.}
	\label{fig3_pesq}
\end{figure}

\begin{figure}[htbp]
	\centering
	\includegraphics[scale=0.5]{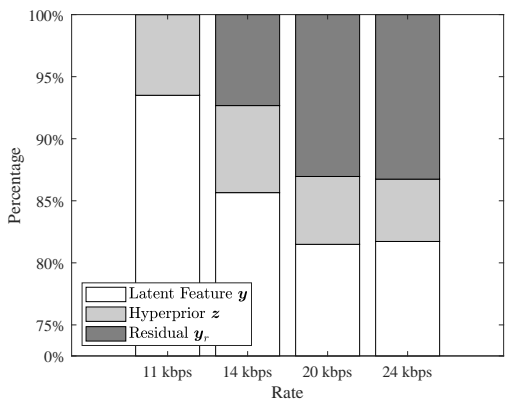}
	\caption{Rate allocation for encoding latent feature $\y$, hyperprior $\z$, and residual latent features $\y_r$ in residual coding scheme.}
	\label{fig4_rate}
\end{figure}

\begin{figure*}[htbp]
    \centering
    \subfloat[12 kbps]{
           \includegraphics[width=0.65\columnwidth]{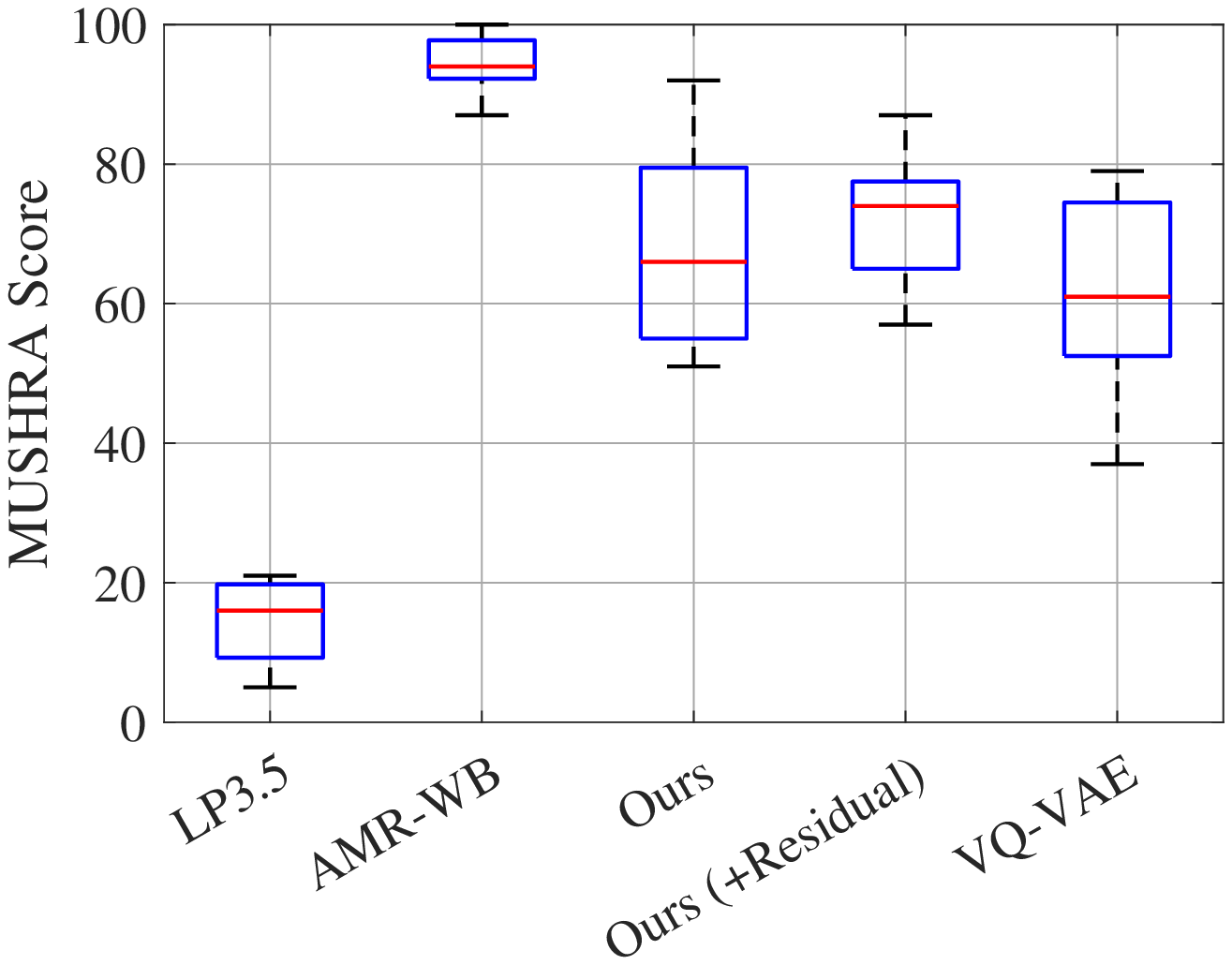}}
    \subfloat[16 kbps]{
           \includegraphics[width=0.65\columnwidth]{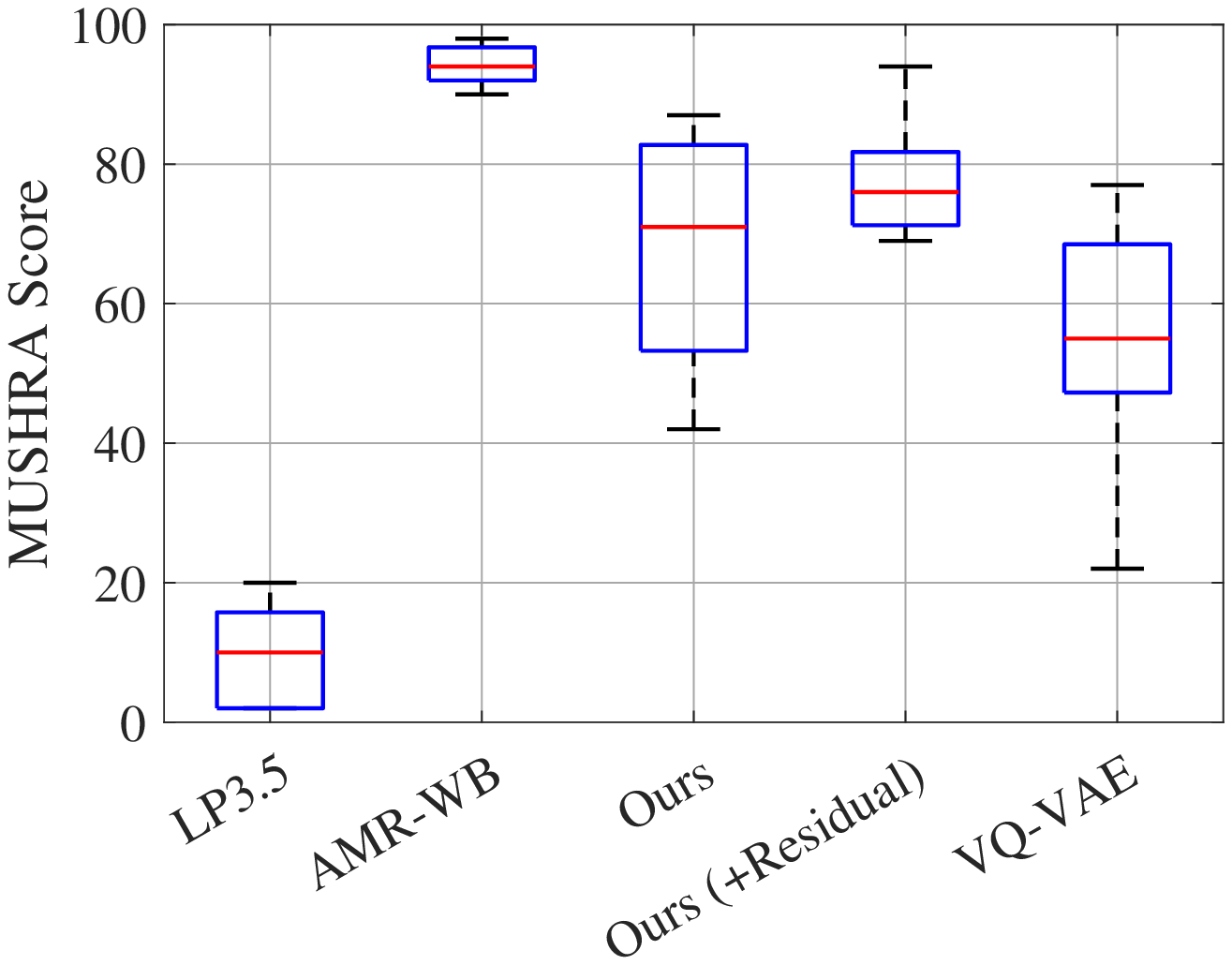}}
    \subfloat[24 kbps]{
           \includegraphics[width=0.65\columnwidth]{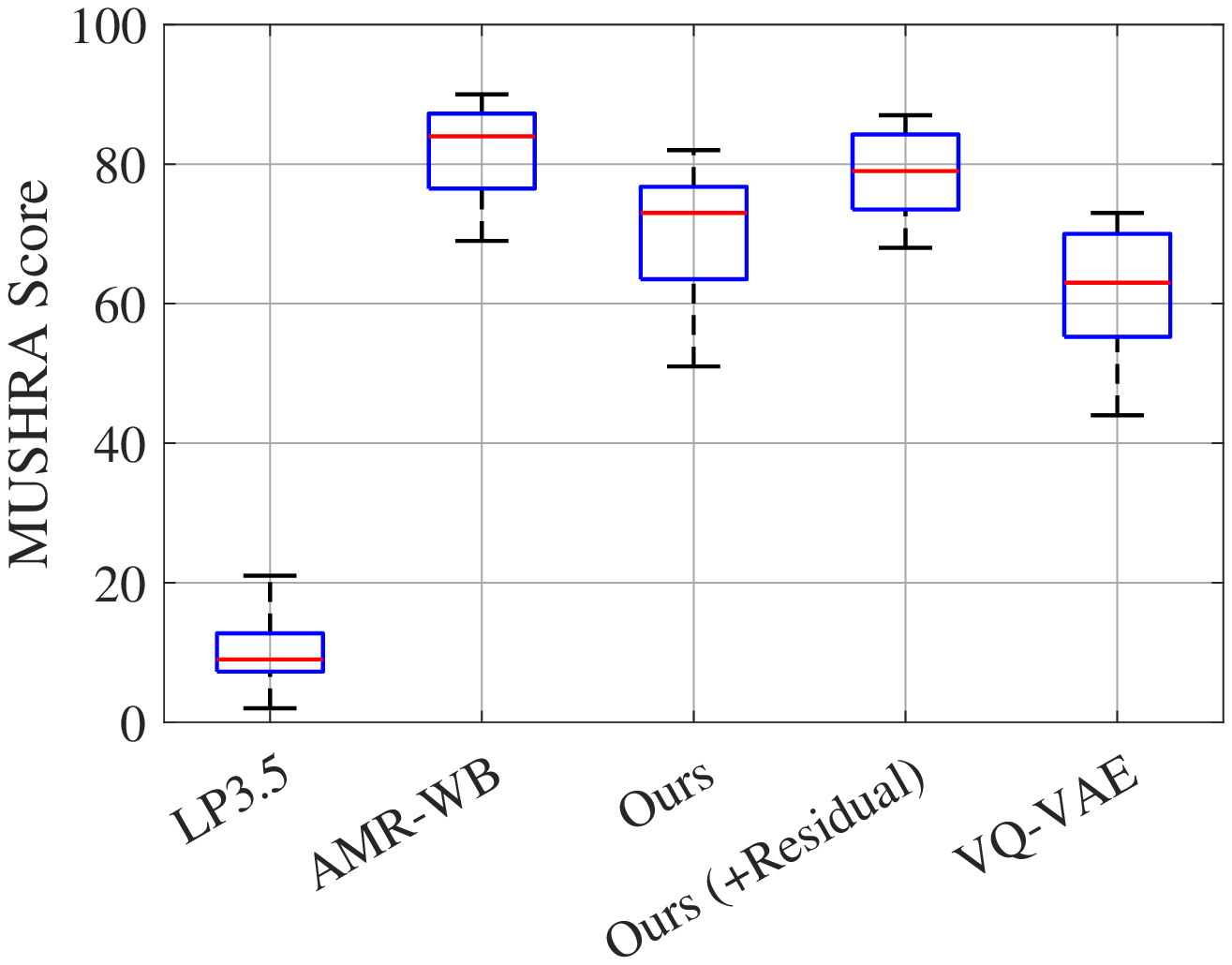}}
    \caption{MUSHRA scores evaluation in three rates, (a) 12 kbps, (b) 16 kbps, (c) 24 kbps. Orange lines are medians. Blue boxes represent the interval between 25th percentile and 75th percentile.}
	  \label{fig5_mushra}
\end{figure*}

In Fig. 3, methods using residual coding are plotted in solid lines, or dashed lines otherwise. Results show that our proposed neural waveform coding achieves a better objective quality than traditional codecs and existing residual waveform coding methods across various coding rates. Our model captures the dependency of the speech features and models its probabilistic distribution more accurately, leading to better RD trade-off. Waveform coding with VQ-VAE shares a similar architecture of our waveform analysis encoder and synthesis decoder, and the posterior categorical distribution is regularized by an entropy constraint. It can be found in the figure that VQ-VAE achieves comparative quality scores with ours at high bitrate with a large codebook to encapsulate the dependency. However, the complexity of searching the codebook becomes high consequently.

In the residual coding settings, LPC-CMRL and CQ compress the LPC coefficients and the LPC residual signal step by step rather than the raw signals. However, LPC residuals depend on the capability of LPC model with quantized coefficients. Our residual coding (solid red line) further improves the reconstruction quality with lower error introduced by the latent quantization. It is observed that our residual coding outperforms the ``Raw-CMRL'' in the objective listening quality, especially in low bitrate region. Besides, in comparison to CMRL, our residual coding does not require auto-regressive encoding. With the same target of objective quality scores, our method can save up to $27\%$ coding rates compared to traditional codecs in low bitrate region.

We also observe the rate allocation in residual coding scheme in Fig. \ref{fig4_rate}. There is a marked upward trend in the cost for encoding the residual latent with the target rate increasing, but the percentage in the total budget is low ($< 15\%$). While in low rate region, the joint training induces the model not to encode the residual latent. It indicates that the performance gain from encoding the primary part of the latent features is superior to that from encoding the residual with the same number of bits.

\subsection{Subjective Tests}
We implement MUSHRA subjective test \cite{mushra}, which is a multi-stimulus method for evaluating medium and large impairments of audio. We randomly select 10 reconstructed waveform signals from the TIMIT test dataset. In Fig. \ref{fig5_mushra}, we compare the subjective performance with a traditional audio codec (AMR-WB) and a neural waveform coder with VQ-VAE defined above. The anchor signal is given by a low-pass filtered signal with cutoff rate at 3.5 kHz (``LP3.5'' in Fig. \ref{fig5_mushra}).

High-pitched artifacts occur in VQ-VAE samples, which are not favored in the listening test. In high bitrate region, our residual coding scheme shows substantial gain over that does not encode residual latent features. However, the reconstructed speech signal is still slightly unsmooth, and noise occurs when it is expected silent. A reasonable explanation is that our waveform coding has not considered the correlation across frames, while linear prediction across frames is adopted in AMR-WB. And the ANNs process the normalized waveforms, and thus the distortion on the points with small magnitude is amplified, according to the nature of human perception.

\subsection{Complexity Analysis}
Table \ref{table_complexity} compares the model size of the proposed speech waveform codec and other neural waveform codecs. For fair comparison, the two-stage models of CMRL and CQ are listed in the table. As we shall note, as an auxiliary coding, the residual coding further alleviates the degradation arising from the quantization of speech latent features with a low extra cost of the model complexity.

\begin{table}[htbp]
\caption{Model parameter comparison of neural waveform coding.}
\renewcommand{\arraystretch}{1.3}
\begin{center}
\begin{tabular}{|c|c|c|c|c|}
\hline
\textbf{Model} & \textbf{Params}($\times10^6$)  \tabularnewline
\hline
CMRL \cite{zhen2019cascaded} & 0.93 (two-stage)  \tabularnewline
\hline
CQ \cite{zhen2020efficient} & 1.35 (two-stage)  \tabularnewline
\hline
VQ-VAE  & 1.57  \tabularnewline
\hline
Ours & 2.31  \tabularnewline
\hline
Ours (+Residual) & 2.57 \tabularnewline
\hline
SoundStream \cite{zeghidour2021soundstream} & 8.40 \tabularnewline
\hline
\end{tabular}
\label{table_complexity}
\end{center}
\end{table}

We additionally verify the superiority regarding the complexity of training our model compared to training VQ-VAE. Our model is trained with 200k steps, and VQ-VAE requires $1.5\times$ training steps at 20 kbps and $2.0\times$ at 24 kbps. As the rate increases, the codebook size of VQ increases, leading to high complexity of searching the codebook in a high-dimensional latent space.

\section{Conclusion}
\label{sec:conclusion}
We have presented a novel neural speech waveform compression method that catalyzes speech semantic communications. The method captures the probabilistic distribution of latent speech features accurately by building an entropy model with hyperprior. It attains a flexible rate-distortion trade-off and the waveform fidelity is optimized for waveform distortion and perceptual distortion. An auxiliary branch is established to encode the residual latent features, improving the speech quality further. Results indicate that the proposed method achieves a better RD performance at various bitrates, and the residual coding scheme outperforms existing residual coding methods, which adopt multi-stage autoregressive coding. Future research may include modeling the correlation across frames in time and frequency domain.

\bibliography{bliography}
\bibliographystyle{IEEEtran}

\vspace{12pt}

\end{document}